\documentstyle[12pt,psfig,epsfig]{article}
\title{\hspace*{10cm} {\large Budker INP 2000-22} \\
\vspace{0.5cm} 
 {\bf Photon colliders: key problems, new ideas}\thanks{Talk at 
the 3nd Int.Workshop on Electron--Electron Interaction at TeV Energies,
 Santa Cruz, CA, USA, December 10--12, 1999. To be published in Int. J. Mod.
Phys. A}}
\author{Valery Telnov\thanks{Email: telnov@inp.nsk.su} \\
{\it Institute of Nuclear Physics,
630090, Novosibirsk, Russia}  \\ 
{\it and DESY, Hamburg, Germany }} 
\date{}

\begin{document}
\maketitle
\newcommand{\EP}{\mbox{e$^+$}}
\newcommand{\EM}{\mbox{e$^-$}}
\newcommand{\EPEM}{\mbox{e$^+$e$^-$}}
\newcommand{\EMEM}{\mbox{e$^-$e$^-$}}
\newcommand{\EE}{\mbox{ee}}
\newcommand{\GG}{\mbox{$\gamma\gamma$}}
\newcommand{\GP}{\mbox{$\gamma$e$^+$}}
\newcommand{\GE}{\mbox{$\gamma$e}}
\newcommand{\LGE}{\mbox{$L_{\GE}$}}
\newcommand{\LGG}{\mbox{$L_{\GG}$}}
\newcommand{\LEE}{\mbox{$L_{\EE}$}}
\newcommand{\TEV}{\mbox{TeV}}
\newcommand{\WGG}{\mbox{$W_{\gamma\gamma}$}}
\newcommand{\GEV}{\mbox{GeV}}
\newcommand{\EV}{\mbox{eV}}
\newcommand{\CM}{\mbox{cm}}
\newcommand{\M}{\mbox{m}}
\newcommand{\MM}{\mbox{mm}}
\newcommand{\NM}{\mbox{nm}}
\newcommand{\MKM}{\mbox{$\mu$m}}
\newcommand{\E}{\mbox{$\epsilon$}}
\newcommand{\EN}{\mbox{$\epsilon_n$}}
\newcommand{\EI}{\mbox{$\epsilon_i$}}
\newcommand{\ENI}{\mbox{$\epsilon_{ni}$}}
\newcommand{\ENX}{\mbox{$\epsilon_{nx}$}}
\newcommand{\ENY}{\mbox{$\epsilon_{ny}$}}
\newcommand{\EX}{\mbox{$\epsilon_x$}}
\newcommand{\EY}{\mbox{$\epsilon_y$}}
\newcommand{\SEC}{\mbox{s}}
\newcommand{\CMS}{\mbox{cm$^{-2}$s$^{-1}$}}
\newcommand{\MRAD}{\mbox{mrad}}
\newcommand{\IND}{\hspace*{\parindent}}
\newcommand{\beq}{\begin{equation}}
\newcommand{\eeq}{\end{equation}}
\newcommand{\beqn}{\begin{eqnarray}}
\newcommand{\eeqn}{\end{eqnarray}}
\newcommand{\dst}{\displaystyle}
\newcommand{\bm}{\boldmath}
\newcommand{\BX}{\mbox{$\beta_x$}}
\newcommand{\BY}{\mbox{$\beta_y$}}
\newcommand{\BI}{\mbox{$\beta_i$}}
\newcommand{\SX}{\mbox{$\sigma_x$}}
\newcommand{\SY}{\mbox{$\sigma_y$}}
\newcommand{\SZ}{\mbox{$\sigma_z$}}
\newcommand{\SI}{\mbox{$\sigma_i$}}
\newcommand{\SIP}{\mbox{$\sigma_i^{\prime}$}}
\newcommand{\n}{\mbox{$n_f$}}

\maketitle

\begin{abstract} 

High energy photon colliders based on laser backscattering
are a very natural extension of a \EPEM\ linear colliders and open new
possibilities to study of the matter. This option has been included in
the pre-conceptual designs of linear colliders and work on Technical
Design Reports is in progress.  The physics motivation for photon
colliders is quite clear though  more studies are needed.  The proof of its
technical feasibility and the search for the best solutions is of
first priority now. In this talk we discuss: physics motivation, laser
problems and new possible solutions, and generation of low emittance beams
needed for obtaining very high luminosities.

\end{abstract}

\section{Introduction.}

  Linear colliders (LC) in the range of a few hundred GeV to several
TeV are under intense study in the world~\cite{NLC}-\cite{JLC}. In
addition to \EPEM\ collisions, linear colliders provide a unique
possibility to study \GG\ and \GE\ interactions at energies and
luminosities comparable to those in \EPEM\
collisions~\cite{GKST81}-\cite{GKST84}. High energy photons can be
produced using laser backscattering. This option has been included in
the conceptual designs of the linear collider
projects and the next goals are the Technical
Design Reports which should be prepared within the next 1--2 years.

\vspace{2mm}

In this short time period we should clearly show that \GG,\GE\
collisions can give new physics information in addition to \EPEM\
collisions and that all technical problems of photon colliders have
solutions. A key element in the project is a powerful laser
system. Solution of this problem is vital for photon colliders.  From the
accelerator side polarized electron beams with very low emittances are
required, smaller than those necessary for \EPEM\ collisions (especially
in the horizontal direction).  At the beginning one can use the same
electron beams as for \EPEM, which is already acceptable for study of
a good physics, but the luminosity in this case will be much lower 
than its real limit determined by collisions effects. In addition,
there are many other technical aspects which should be developed and
taken into account in the basic LC designs (before beginning of the
construction).

\vspace{2mm} 

In this paper we will briefly consider physics motivation, new ideas on 
laser schemes, ways to high luminosities and associated problems.

\section{Physics}

Physic motivation of photon colliders is quite clear. In general, the
physics in \GG,\GE\ collisions is complimentary to that in \EPEM\
interactions. Some phenomena can best be studied at photon
colliders. Several examples are given below. 

The present ``Standard'' model predicts a very unique particle, the
{\it Higgs boson}.  It is very likely that its mass is about 100--200 GeV,
i.e., lays in the region of the next linear colliders.
In \GG\ collisions the Higgs boson will be produced as a single
resonance. This process goes via the loop and its cross section is
very sensitive to all heavy (even super-heavy) charged particles.  The
graphs for effective cross section of the Higgs production in \GG\
collisions and in \EPEM\ collisions can be found
elsewhere~\cite{ee97}. For $M_H=$ 120--250 GeV the effective cross
section \GG\ in collisions is larger than that in \EPEM\ collisions by a
factor of about 6--30.  The Higgs can be detected as a peak in the invariant
mass distribution or can be searched for by energy scanning using the
very sharp high energy edge of luminosity distribution~\cite{ee97}.  

The cross section of the process $\gamma\gamma \to H \to b\bar{b}$ is
proportional to $\Gamma_{\GG}(H)\times Br(H\to b\bar{b}$). The
branching ratio $Br(H\to b\bar{b})$ can be measured with high
precision in \EPEM\ collisions~\cite{Bataglia}. As a result, one can
measure the $\Gamma_{\GG}(H)$ width at photon colliders with an
accuracy better than 2-3\%~\cite{Jikia},\cite{Melles}. The value of
the two-photon decay width is determined by the sum of contributions
to the loop of all heavy charge particles with masses up to infinity.
So, it is a unique way to ``see'' particles which cannot be produced at
the accelerators directly (maybe never).

The Higgs two-gluon decay width (which can be extracted from joint LHC, LC
data) is also sensitive to heavy particles in the loop, but only to
those which have strong interactions.  These two
measurements together with the $\Gamma_{Z\gamma}(H)$ width, which
could be measured in \GE\ collisions, will allow us to ``observe'' and
perhaps understand the nature of invisible heavy charged
particles. This would be a great step forward in our understanding of
the matter.

The second example is the {\it charged pair production}.  Cross
sections for the production of charged pairs in \GG\ collisions are
larger than those in \EPEM\ collisions by a factor of approximately
5--10~\cite{TEL95},\cite{TESLA}. The cross section of the scalar
pair production in collisions of polarized photons near the threshold,
is higher than that in \EPEM\ collisions by a factor of 10--20 (see
figures in Refs.~\cite{TKEK},\cite{Tfrei}).  Near the threshold the
cross section in the \GG\ collisions is very sharp (while in \EPEM\ it
contains a factor $\beta^3$) and can be used for measurement of
particle masses.
Note, that in \EPEM\ collisions two charged pairs are produced both via
annihilation diagrams with virtual $\gamma$ and $Z$ and also via
exchange diagrams where new particles can contribute, while in \GG\
collisions it is pure QED process which allows the spin and charge of
produced particles to be measured unambiguously.  

In \GE\ collisions, charged particles with a mass higher than that in
\EPEM\ collisions can be produced (a heavy charged particle plus a
light neutral), for example, supersymmetric charged particle plus
neutralino or new W boson and neutrino.  \GG\ collisions also provide
higher accessible masses for particles which are produced as a single
resonance in \GG\ collisions (such as the Higgs boson).

A new theory ``Quantum gravity effects in Extra Dimensions'' proposed
recently~\cite{Arkani} suggests a possible explanation of the fact
why gravitation forces are so weak in comparison with electroweak
forces.  It turns out that this extravagant theory can be tested at
linear colliders, moreover, photon colliders are sensitive up to a
factor of 2 higher quantum gravity mass scale than \EPEM\
collisions~\cite{RIZZO}.

\section{Lasers, Optics}

The new key element at photon colliders is a powerful laser system
which is used for e$\to \gamma$ conversion.  The laser system should
have the following parameters: flash energy of 2-5 J with
``diffraction'' quality, wave length about 1 \MKM\ (for 2E = 500 GeV),
pulse duration about 1 ps, repetition rate  10--15 kHz with
the same pulse structure as for electron beams. Detailed
consideration of requirements to lasers can be found
elsewhere~\cite{TEL90},\cite{TEL95},\cite{Monter},
\cite{TESLA},\cite{Tfrei},\cite{Tsit1},\cite{Las1}.

Picosecond Terawatt laser pulses can be produced using the chirped
pulse technique which allowed a peak laser power to be increased by
three order of magnitude during the last decade. The main problem at
photon colliders is the high repetition rate and correspondingly very
high average power, about 50 kW. An additional complication is due to the
train structure of pulses at LC which means even larger average power
inside the train.  One very promising way to overcome this problem is
discussed in this paper.  It is an optical cavity approach, which
allows a considerable reduction of the required peak and average laser
power.

\subsection{One-pass Laser Systems}

Let me start with a discussion of a one pass laser system. One possible
solution is the multi-laser system where  pulses are combined into one
train using Pockels cells~\cite{NLC}. Several (about 20 in the NLC Zero
Design Project) lasers allow  spliting the average power between  lasers
and thus solving the problem with cooling of the amplifying medium.  It is
clear that these lasers should use crystals with high thermal
conductivity, diode (semiconductor lasers) pumping with very high
efficiency, adaptive optics. These topics were discussed at this workshop
in a nice talk given by M.~Perry.  He pointed out two main problems for NLC
laser system (these problems are not essential for TESLA):

1. Due to the short distance between electron bunches (2.8 ns between
bunches, with 95 bunches in the train), the power of the diode pumping
system should be very high, about 2 J/2.8 ns/efficiency, that is
huge. The estimated cost of the diode  system is about 100 M\$.

2. Combining Pockels cells should be very fast (switching time is less than
3 ns) which is very difficult (or even impossible) for the considered powers.

    In my opinion, the situation here is not so bad and there are ways 
to overcome these problems. 
\begin{figure}[!htb]
\centering
\vspace*{0.2cm} 
\hspace*{-0.2cm} \epsfig{file=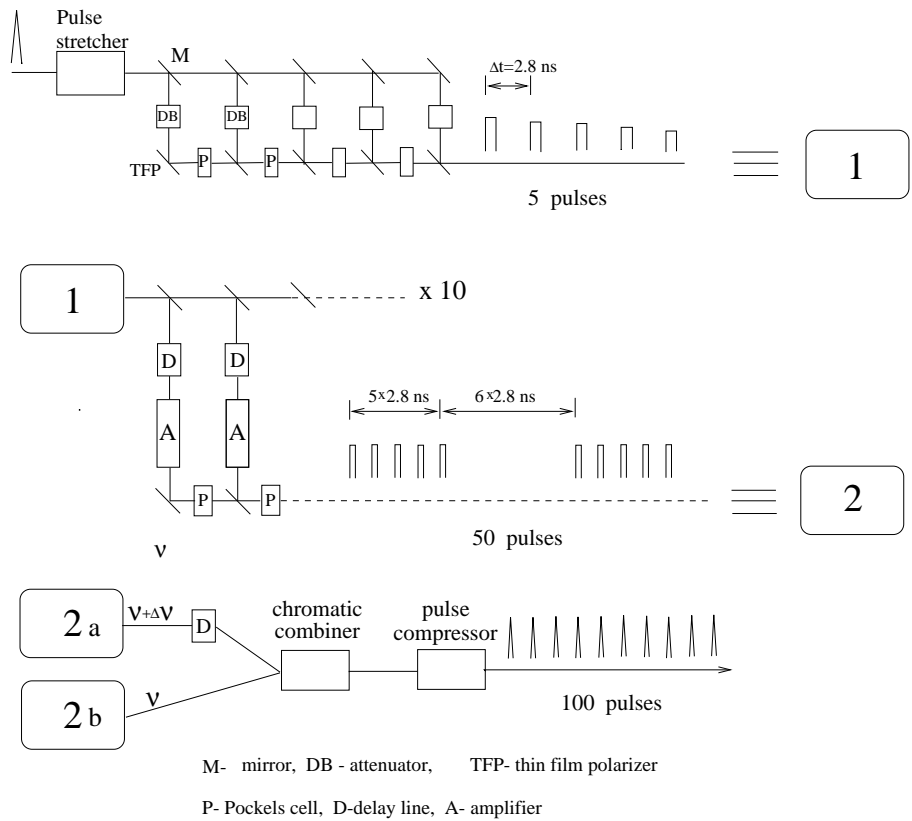,width=12cm,angle=0} 
\vspace*{0.2cm} 
\caption{Possible laser scheme  for one-pass mode of operation at NLC/JLC}
\vspace{0mm}
\label{onepass}
\vspace{-2mm}
\end{figure} 
First, the storage time of most promissing crystals such as Yb:S-FAP
is about 1 msec, so I do not understand why diodes should work only
during 270 ns (length of the pulse train). The ratio of these times is 3700!
These crystals can be pumped before the pulse train during more than 2
orders longer time! What is wrong?  Discharge by spontaneous radiation
can be suppresed by proper choice of geometry or (and) use of optical
locks. Pulse energies in several sequential pulses passing one
amplifier can be equalized by adjusting the pulse energies of incoming
pulses. If all these is correct, one can decrease the pumping power at
least by 2 orders of magnitude and there will be no problem.

The second problem also has a solution, see Fig.~\ref{onepass}. Using
Pockels cells we can prepare two trains of laser pulses with somewhat
different average frequences ($\Delta \nu \sim$ bandwidth) and then
join them using a chromatic combiner. Each of the trains is prepared in
the following way.\footnote{The technique of combining pulses in the
train using Pockels cells and thin-film-polarizers is explained
elsewhere~\cite{NLC}.} At the beginning (before the final amplifiers)
we manipulate with low energy pulses and prepare 10 short subtrains
consisting of 5 pulses spaced by 2.8 ns. At this stage, power is low
and there is no problem with Pockels cells. Then these subtrains are
amplifiered and combined using Pockels cells into one train with the
distance between subtrains equal to the length of one subtrain which
is equal 6$\times$2.8 ns.  Here we can use already rather slow Pockels
cells with larger diameter, so that the power/cm$^2$ can be much lower
than that for 2.8 ns Pockels cells.

So, it seems, both problems have solutions. Of course, a final answer
should be given by laser experts.

\subsection{Multi-pass Laser Systems}

To overcome the ``repetition rate'' problem in a radical way it is quite
natural to consider a laser system where one laser bunch is used for
e$\to \gamma$ conversion many times. Indeed, one Joule laser flash
contains about $10^{19}$ laser photons and only $10^{10}-10^{11}$
photons are knocked out in the collision with one electron bunch.

The simplest solution is to trap the laser pulse to some optical loop and
use it many times~\cite{NLC}. In such a system the laser pulse enters
via the film polarizer and then is trapped using Pockels cells and
polarization rotating plates.  Unfortunately, such a system will not
work with Terawatt laser pulses due to a self-focusing effect.

However, there is one way to ``create'' a powerful laser pulse in
the optical ``trap'' without any material inside. This very promising
technique is discussed below.	
Shortly, the method is the following. Using the train of low energy
laser pulses one can create in the external passive cavity
(with one mirror having some small transparency) an optical pulse of
the same duration but with much higher energy (pulse stacking). This
pulse circulates many times in the cavity  each time colliding with
electron bunches passing the center of the cavity.

The idea of pulse stacking is simple but not trivial and not well
known in the HEP community. This method is used now in several
experiments on detection of gravitation waves. It was mentioned also
in NLC ZDR~\cite{NLC}, though without analysis and further development.
 In my opinion, pulse stacking is very natural for photon colliders
and allows not only  building a relatively cheap laser system for
$e\to\gamma$ conversion, but gives us the practical way for realization of the
laser cooling, i.e. opens up the way to ultimate luminosities of photon
colliders. 
\begin{figure}[!htb]
\centering
\vspace*{0.2cm} 
\hspace*{-0.2cm} \epsfig{file=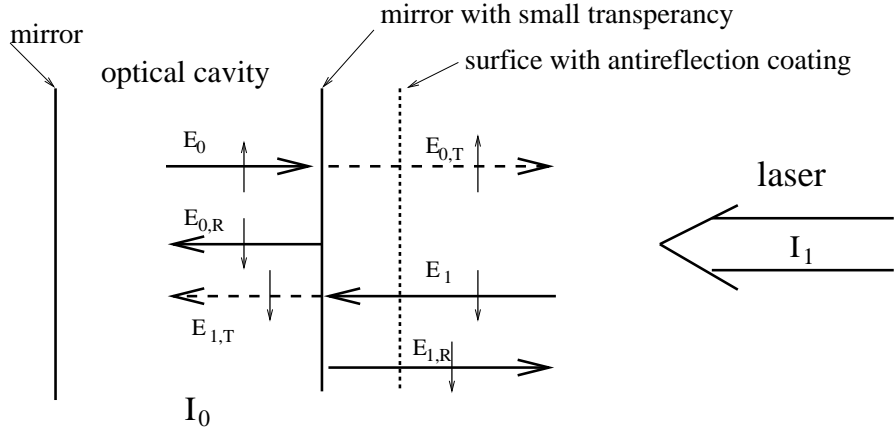,width=12cm,angle=0} 
\vspace*{0.2cm} 
\caption{Principle of pulse stacking in an external optical cavity.}
\vspace{2mm}
\label{cavity}
\vspace{-5mm}
\end{figure} 

As this is very important for photon colliders, let me consider this
method in more detail~\cite{Tfrei}. The principle of pulse stacking is
shown in Fig.~\ref{cavity}.
The secret consists of the following. There is a well known optical
theorem: at any surface, the reflection coefficients for light coming
from one and the other sides have opposite signs. In our case, this
means that light from the laser entering through a semi-transparent
mirror into the cavity interferes with reflected light inside the
cavity {\it constructively}, while the light leaking from the cavity
interferes with the laser light reflected from the cavity {\it
destructively}. Namely, this fact produces asymmetry between cavity
and space outside the cavity!

Let R be the reflection coefficient, T the transparency coefficient
and $\delta$ the passive losses in the right mirror. From the energy
conservation $R+T+\delta =1$. Let $E_1$ and $E_0$ be the amplitudes
of the laser field and the field inside the cavity. In equilibrium,
$E_0= E_{0,R} + E_{1,T}$. 
Taking into account that $E_{0,R}=E_0\sqrt{R}$, $E_{1,T}=E_1\sqrt{T}$ and 
$\sqrt{R}\sim 1-T/2-\delta/2$ for $R\approx 1$, we obtain
$E_0^2/E_1^2=4T/(T+\delta)^2.$ The maximum ratio of intensities 
is obtained at $T=\delta$, then $I_0/I_1=1/\delta \approx Q$,
where $Q$ is the quality factor of the optical cavity.  Even with two
metal mirrors inside the cavity, one can hope to get a gain factor of about
50--100; with multi-layer mirrors it can reach $10^5$. The TESLA
collider has 2800 electron bunches in the train, so a factor of
1000 would be perfect for our goal, but even a factor of ten
means a drastic reduction of the cost.

   Obtaining high gains requires a very good stabilization of cavity
size: $\delta L \sim \lambda/4\pi Q$, laser wave length: $\delta
\lambda/\lambda \sim \lambda/4\pi QL$, and distance between the laser
and the cavity: $\delta s \sim\lambda/4\pi$. Otherwise, the  condition of
constructive interference will  not be fulfilled. Besides, the
frequency spectrum of the laser should coincide with the cavity modes,
that is automatically fulfilled when the ratio of the cavity length and
that of the laser oscillator is equal to an integer number 1, 2, 3... . 

      In HEP literature I have found only one reference on pulse
stacking of short pulses ($\sim 1$ ps) generated by the free electron
laser with the wave length of 5 $\mu$m~\cite{HAAR}. They observed
pulses in the cavity with 70 times the energy of the incident FEL
pulses, though no long term stabilization was done.

\begin{figure}[!htb]
\centering
\vspace*{-0.5cm} 
\hspace*{-0.4cm} \epsfig{file=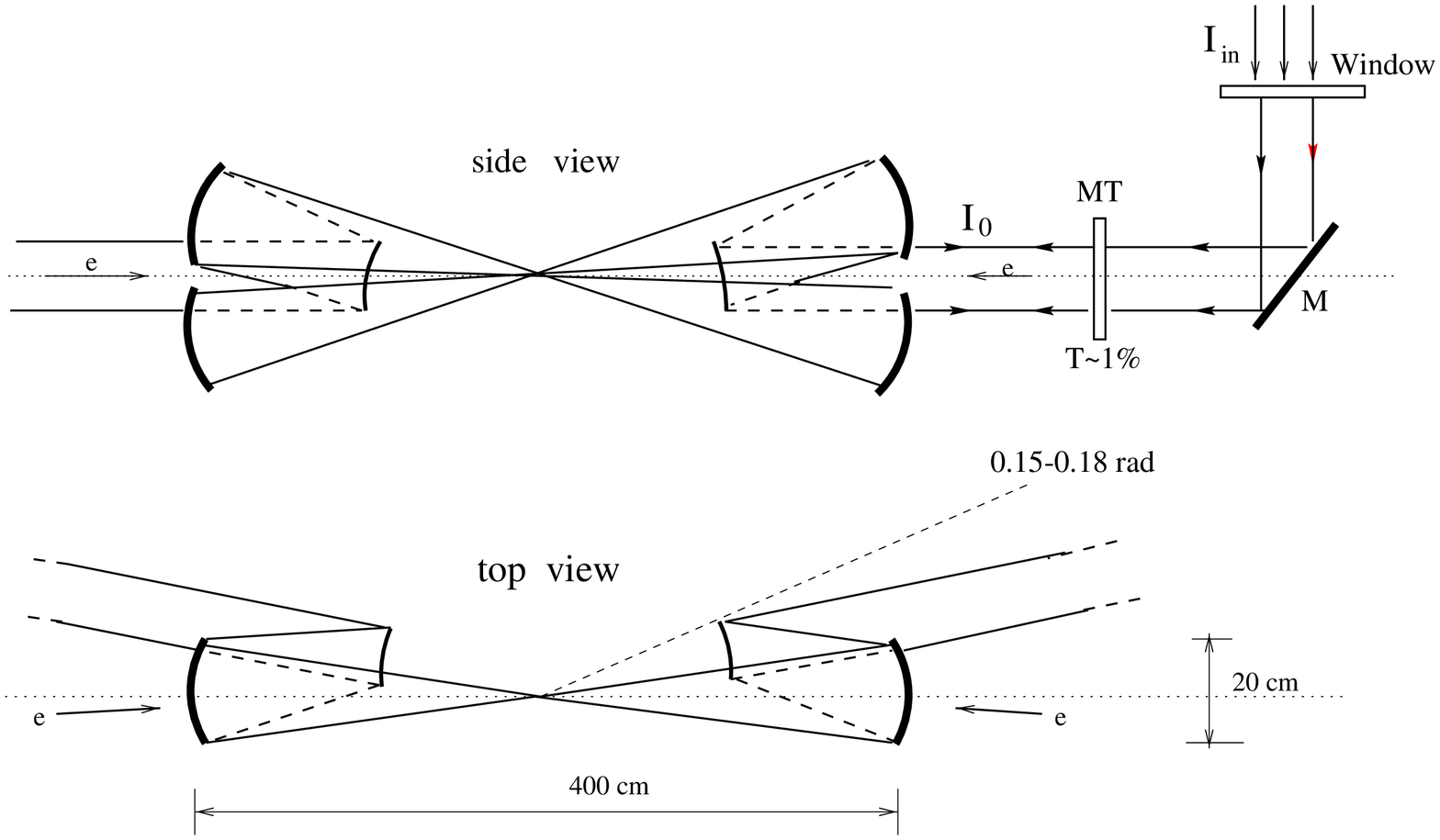,width=13cm,angle=0} 
\vspace*{0.4cm} 
\caption{Possible scheme of optics at the IR.}
\vspace{0mm}
\label{optics}
\vspace{-2mm}
\end{figure} 

Possible layout of the optics at the interaction region scheme is
shown in Fig.~\ref{optics}. There are two optical cavities 
(one for each colliding electron beam) placed outside the
electron beams. The required flash energy in this case is larger by a
factor of 2 than in the case of head-on collisions but all other problems
(holes in mirrors etc) are much simpler.

\section{Luminosity of Photon Colliders in Current Designs}
Some results of simulation of \GG\ collisions at TESLA, ILC (converged
NLC and JLC) are presented below in Table \ref{table1}. Beam
parameters were taken the same as those in \EPEM\ collisions with the
exception of the horizontal beta function at the IP which is taken
(quite conservatively) equal to 2 mm for all cases, that is several
times smaller than that in \EPEM\ collisions.  The conversion point
(CP) is situated at distance $b=\gamma\sigma_y$. It is assumed that
the ``Compton'' parameter $x=4.6$, electron beams have 85\% longitudinal
polarization and laser photons have 100\% circular polarization.

\begin{table}[hbt]
\vspace{.1cm}
\caption{Parameters of  \GG\ colliders based on TESLA, ILC (NLC/JLC).}
\vspace{.1cm}
\renewcommand{\arraystretch}{1}
\begin{center}
\hspace*{-2.3mm}\begin{tabular}{l c c c c } \hline
& T(500) & I(500)   &  T(800) & I(1000)  \\ \hline  
$N/10^{10}$& 2. & 0.95 & 1.4 & 0.95 \\  
$\sigma_{z}$, mm & 0.4 & 0.12 & 0.3 & 0.12 \\  
$f_{rep}\times n_b$,& 5$\times$2821 & 120$\times$ 95& 3$\times$4500 & 120$
\times$95 \\
$\Delta t_b$, ns  & 337 & 2.8 & 189 & 2.8  \\ 
$\gamma \epsilon_{x,y}/10^{-6}$,m$\cdot$rad & $10/0.03$ & $5/0.1$ 
&  $8/0.01$ & $5/0.1$ \\
$\beta_{x,y}$,mm at IP& $2/0.4$ & $2/0.12$  &
$2/0.3$& $2/0.16$  \\
$\sigma_{x,y}$,nm& $200/5$ & $140/5$ & 
$140/2$ & $100/4$ \\  
$L(geom),\,\,\,  10^{33}$& 48 & 12 & 75 & 20 \\  
$\LGG (z>0.65), 10^{33} $ & 4.5 & 1.1  & 7.2 & 1.75  \\
$\LGE (z>0.65), 10^{33}$ & 6.6 & 2.6 & 8  & 4.2  \\
$\LEE, 10^{33}$ & 1.2 & 1.2  & 1.1 & 1.8 \\ \hline
\vspace{-5.mm}
\end{tabular}
\end{center}
\label{table1}
\end{table}

The \GG\ luminosity in these projects is determined only by 
``geometric'' ee-luminosity. With some new low
emittance electron sources  one can get, in principle,
$\LGG (z>0.65) > L_{\EPEM}$.  The limitations and technical
feasibility are discussed in the next section.  
  
\begin{figure}[!htb]
\centering
\vspace*{-1.5cm} 
\hspace*{-0.6cm} \epsfig{file=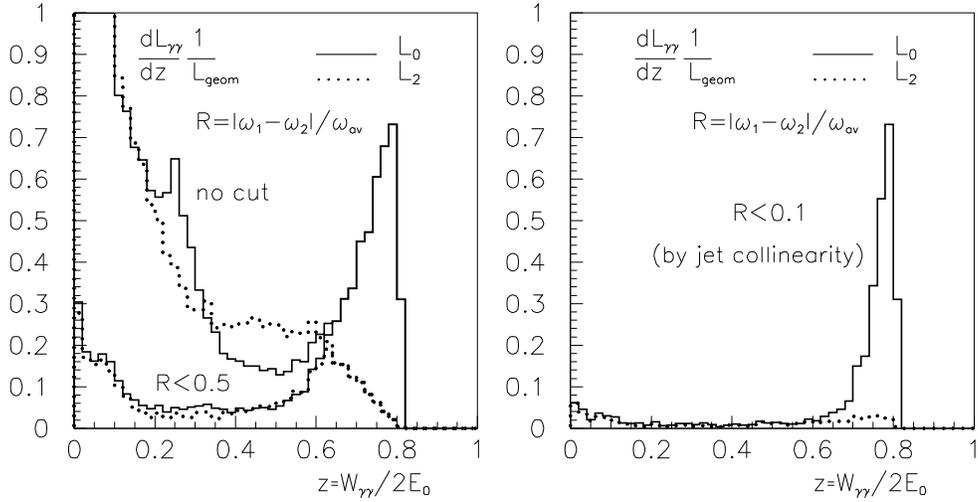,width=16.5cm,angle=0} 
\vspace{-2.3cm} 
\caption{\GG\ luminosity spectra at TESLA(500) for parameters
presented in Table 1. See comments in the text.}
\vspace{-0.2cm} 
\label{TeslaR}
\end{figure}

   The normalized \GG\ luminosity spectra for a 0.5 TeV TESLA are
shown in Fig.~\ref{TeslaR}(left).
We see that, in the high energy part
of the luminosity spectra, photons have a high degree of polarization,
which is very important for many experiments.  In addition to the high
energy peak, there is a factor 5--8 larger low energy luminosity.
 The  events in this region have a large boost and
can be easily distinguished from the central high energy events.  In
the same Fig.~\ref{TeslaR} (left) you can see the same spectrum with an
additional ``soft'' cut on the longitudinal momentum of the produced
system which suppresses low energy luminosity to a negligible level.

Fig.~\ref{TeslaR} (right) shows the same spectrum with a stronger cut
on the longitudinal momentum. In this case, the spectrum has a nice
peak with the width at half of maximum about 7.5\%. For two jet
events one can obtain this nice ``collider resolution'' without
accurate energy measurement in the detector applying only a cut on the
acollinearity angle between jets ($H\to b\bar b, \tau\tau$, for
example).

 A  similar table and distributions for the photon collider on the
c.m.s. energy 130 GeV (Higgs collider) can be found elsewhere~\cite{TKEK}.
\begin{figure}[!htb]
\centering
\vspace*{-0.9cm} 
\hspace*{-0.7cm} \epsfig{file=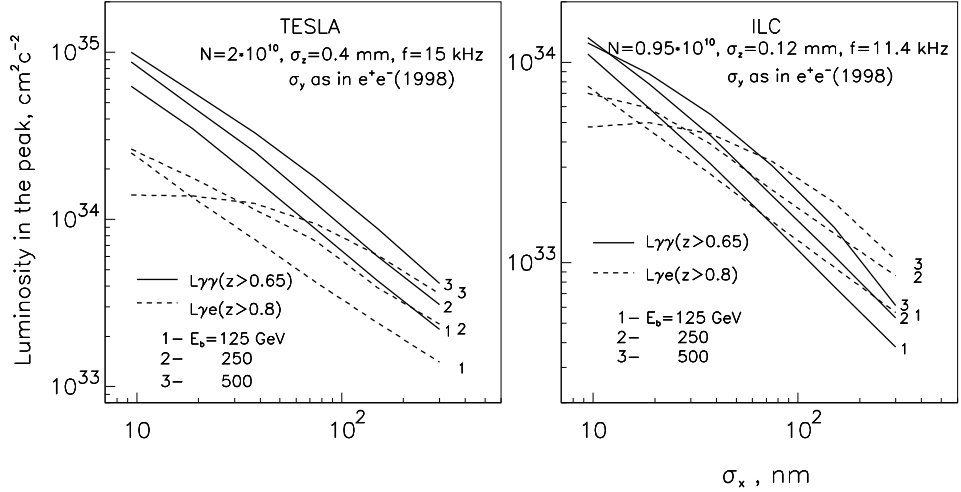,width=14cm,angle=0} 
\vspace*{-0.9cm} 
\caption{Dependence of \GG\ and \GE\ luminosities in the high energy
peak on the horizontal beam size for TESLA and ILC at various
energies. See also comments in the text.}
\vspace{-0.3mm}
\label{sigmax}
\vspace{-0mm}
\end{figure} 

\section{Ultimate {\boldmath \GG, \GE\ } Luminosities }
In the current projects the \GG\ luminosities are determined by the
``geometric'' luminosity of the electron beams. Having electron beams
with smaller emittances one can obtain a much higher \GG\ 
luminosity~\cite{TSB2}. Below are results of the simulation with the
code which takes into account all main processes in beam-beam
interactions~\cite{TEL95}. Fig.~\ref{sigmax} shows dependence of the
\GG\ (solid curves) and \GE\ (dashed curves) luminosities on the
horizontal beam size.
 The vertical emittance is taken as in TESLA(500), ILC(500)
projects (see Table \ref{table1}). The horizontal beam size was varied
by changing  horizontal beam emittance keeping the horizontal beta
function at the IP equal to 2 mm.
One can see that all curves for \GG\ luminosity follow their natural
behavior: $\L\propto 1/\sigma_x$, with the exception of ILC at
$2E_0=1$ GeV where at small $\sigma_x$ the effect of coherent pair
creation~\cite{TEL90} is seen. This means that at the same collider
the \GG\ luminosity can be increased by decreasing the horizontal beam
size (see Table 1) at least by one order ($\sigma_x < 10$ nm is
difficult due to some effects connected with the crab crossing).
Additional increase of \GG\ luminosity by a factor about 3 (TESLA),
7(ILC) can be obtained by a further decrease of the vertical
emittance~\cite{TKEK}. So, using beams with smaller emittances, the
\GG\ luminosity at TESLA, ILC can be increased by almost 2 orders of
magnitude. However, even with one order improvement, the number of
``interesting'' events (the Higgs, charged pairs) at photon colliders
will be larger than that in \EPEM\ collisions by about one order. This
is a nice goal and motivation for photon colliders.

  In \GE\ collision (Fig. \ref{sigmax}, dashed curves), the behavior of
the luminosity on $\sigma_x$ is different due to additional collision
effects: beam repulsion and beamstrahlung. As a result, the
luminosity in the high energy peak is not proportional to the
``geometric''  luminosity.

There are several ways of decreasing the transverse beam emittances
(their product): optimization of storage rings with long wigglers,
development of low-emittance RF (or pulsed photo-guns) with merging
many beams with low charge and emittance~\cite{ee97}.  Here some progress is
certainly possible.  Moreover, there is one method which allows
further decrease of beam cross sections by two orders of magnitude in
comparison with current designs; it is a laser 
cooling~\cite{TSB1},\cite{Monter},\cite{Las1}.

\section{Conclusion}

The physics program for photon \GG, \GE\ colliders is very interesting
and the additional cost of the second interaction region is certainly
justified.

Special effort is required for the development of the laser and optics
which are the key elements of photon colliders.  The present laser
technology has, in principle, all elements needed for photon
colliders, the development of a practical scheme is the most pressing
task now. Two laser schemes has been discussed in this paper. The optical
cavity approach allows a considerable reduction of the required
peak and average laser power.

The \GG\ luminosity at photon colliders can be higher than that in
\EPEM\ collisions, typical cross sections are also several times
higher, so one could consider an X-factory (X = Higgs, W, etc.). The
main problem here is the generation of polarized electron beams with
very small emittances.  Optimization of damping rings and development
of low emittance multi-gun RF sources is the first step in this
direction.  The second step requires new technologies. The laser
cooling of electron beams is one possible way of achieving ultimate
\GG\ luminosity. Realization of this method depends on the progress of
laser technology; especially promising is the method of laser pulse
stacking pulses in an optical cavity.

section{Acknowledgements}
\noindent
 I would like to thank Clem Heusch and Nora Rogers for organization of
a nice Workshop which was one of the important steps towards 
\EPEM, ee, \GE, \GG\ colliders.

\end{document}